\documentclass[a4paper,11pt]{article}
\usepackage{jheppub} 
\usepackage{orcidlink}

\usepackage{subfigure}
\usepackage{mathtools}
\usepackage{amsmath}
\usepackage{amsfonts}
\usepackage{amssymb}
\usepackage{bbold}
\usepackage{bm}
\usepackage{ esint }
\usepackage{orcidlink}

\newcommand{\br}{{\bf r}}

\newcommand{\bk}{{\bf k}}

\newcommand{\bv}{{\bf v}}

\newcommand{\bp}{{\bf p}}

\newcommand{\tomegaE}{{\tilde{\omega}_{E}}}
\newcommand{\omegaE}{{\omega_{E}}}

\newcommand{\vc}{v_{\rm c}}

\newcommand{\sA}{{\sf A}}

\newcommand{\sH}{{\sf H}}
\newcommand{\sM}{{\sf M}}
\newcommand{\sN}{{\sf N}}

\newcommand{\oP}{{\overline{P}}}

\newcommand{\opsi}{{\overline{\psi}}}

\newcommand{\cV}{c_{\rm V}}
\newcommand{\sV}{s_{\rm V}}
\newcommand{\tV}{\theta_{\rm V}}

\newcommand{\exclude}[1]{{}}
\long\def\exclude#1{}

\newcommand{\GF}{G_{\rm F}}

\title{Theory of neutrino slow flavor evolution. Part~II. Space-time evolution of linear instabilities.}

\author[a]{Damiano F.\ G.\ Fiorillo \orcidlink{0000-0003-4927-9850}} 
\affiliation[a]{Deutsches Elektronen-Synchrotron DESY,
Platanenallee 6, 15738 Zeuthen, Germany}

\author[b]{and Georg G.\ Raffelt
\orcidlink{0000-0002-0199-9560}}
\affiliation[b]{Max-Planck-Institut f\"ur Physik, Boltzmannstr.~8, 85748 Garching, Germany}

\abstract{Slow flavor evolution (defined as driven by neutrino masses and not necessarily ``slow'') is receiving fresh attention in the context of compact astrophysical environments. In Part~I of this series, we have studied the slow-mode dispersion relation following our recently developed analogy to plasma waves. The concept of resonance between flavor waves in the linear regime and propagating neutrinos is the defining feature of this approach. It is best motivated for weak instabilities, which probably is the most relevant regime in self-consistent astrophysical environments because these will try to eliminate the cause of instability. We here go beyond the dispersion relation alone (which by definition applies to infinite media) and consider the group velocities of unstable modes that determines whether the instability relaxes \textit{within} the region where it first appears (absolute), or away from it (convective). We show that all weak instabilities are convective so that their further evolution is not local. Therefore, studying their consequences numerically in small boxes from given initial conditions may not always be appropriate.
}

\begin{document}
\maketitle
\flushbottom

\section{Introduction}\label{sec:introduction}

The flavor evolution of a sufficiently dense neutrino gas, such as in a supernova (SN) core or in a neutron star merger (NSM), proceeds primarily due to the weak interaction among neutrinos~\cite{Pantaleone:1992eq}
because the evolution driven by masses and flavor mixing is suppressed by the large matter effect \cite{Wolfenstein:1979ni}.
The weak potential produced by neutrinos which are not exactly in a flavor eigenstate induces neutrinos to evolve away from their flavor eigenstates and can trigger runaway modes of flavor conversion which are known as flavor instabilities~\cite{Samuel:1993uw, Samuel:1995ri, Duan:2005cp, Duan:2006an}. These forms of collective evolution may be understood in terms of pairwise conversions of the type $\nu_e\overline{\nu}_e\to \nu_x\overline{\nu}_x$ ($x=\mu$ or $\tau$), which can also happen via scattering processes, but are sped up on the refractive level through coherent transformations driven by the weak interaction potential, effectively realizing a collisionless neutrino plasma. 

How these instabilities are triggered and evolve in an astrophysical environment is mostly an open question. The evolution of the runaway modes can be studied in detail in the linear phase, when the deviation from flavor eigenstates is small. The corresponding linear stability analysis (LSA) acts as a diagnostic tool, answering the question of whether an initial flavor configuration remains stable, or becomes unstable by developing exponentially growing flavor conversions. In the latter case, LSA can only describe the early phases of evolution, before the initial flavor perturbation becomes too large. Still, LSA is the main tool to develop intuition, since it helps us understand the conditions under which the system develops runaway modes.

The main insight that, over the years, has been developed from LSA is that a flavor instability always appears, provided that it is allowed by the conservation laws, and specifically the conservation of lepton number \cite{Johns:2024bob, Fiorillo:2024bzm}. This statement has not been formally proven, yet it seems verified by all known forms of flavor conversions. In the limit of vanishing mass splittings, the individual oscillation frequencies of neutrino flavor eigenstates is independent of energy and is equal for neutrinos and antineutrinos, so only the energy-integrated lepton number matters (neutrinos minus antineutrinos), which is often called the $e$--$x$ lepton number. If this quantity changes sign across the angular distribution of the neutrinos (a so-called \textit{angular crossing}), then an instability is guaranteed~\cite{Morinaga:2021vmc, Dasgupta:2021gfs, Fiorillo:2024bzm}, and its growth rate is independent of the vacuum frequency and is completely determined by the refractive energy shift due to neutrinos, which is of the order of $\mu=\sqrt{2}\GF(n_\nu+n_{\overline\nu})$, where $\GF$ is the Fermi constant and $n_\nu$ and $n_{\overline\nu}$ the neutrino and antineutrino number densities. In fact, since only the lepton number can enter the dynamics, by introducing the lepton-over-neutrino number ratio $\epsilon=(n_\nu-n_{\overline{\nu}})/(n_\nu+n_{\overline{\nu}})$, the growth rate of these instability is of the order of $\gamma\sim \mu \epsilon$. Since this rate is orders of magnitude larger than the vacuum frequency, this form of collective transformation is known as \textit{fast flavor instability}~\cite{Sawyer:2004ai, Sawyer:2008zs, Sawyer:2015dsa, Chakraborty:2016lct, Izaguirre:2016gsx, Airen:2018nvp, Johns:2019izj, Padilla-Gay:2021haz,  Fiorillo:2023mze, Fiorillo:2023hlk, Fiorillo:2024qbl, Fiorillo:2024bzm, Fiorillo:2024uki}. The fast-growing modes usually correspond to inhomogeneous waves with a typical wavelength comparable with the inverse growth rate, a conclusion that can be understood intuitively since, over a timescale $\tau\sim (\mu\epsilon)^{-1}$ during which the perturbation  grows, neutrinos stream over comparable distances, spontaneously developing inhomogeneities over such scales.

On the other hand, even without an angular crossing, the system may still be unstable if lepton number changes among neutrinos and antineutrinos, essentially ensuring the presence of $\nu_e$ and $\overline{\nu}_e$ for pairwise conversions. In Part~I of this series~\cite{Fiorillo:2024pns}, we have considered these \textit{slow instabilities}, showing that they usually grow in time with a rate $\gamma\sim \omega_E\cos2\tV/\epsilon$, where $\omega_E=\delta m^2/2E$ is the energy splitting caused by the squared-mass splitting $\delta m^2$ at energy $E$, and $\tV$ is the mixing angle in vacuum. While the typical timescale for instability growth is relatively large, the length scales of the instability are still very short, of the order of $\lambda\sim (\mu\epsilon)^{-1}$, similar to the fast instability. These slow instabilities will grow in time even in the inner regions of a SN, provided that $\gamma$ exceeds the collisional rate of neutrinos. 
The results of Part~I suggest that a complete focus on fast flavor instabilities may be missing part of the point, since slow instabilities can also affect the properties of the SN core, even in the absence of angular crossings, as also noticed earlier in numerical examples in Ref.~\cite{DedinNeto:2023ykt}.

Our conclusions in Part~I show that slow instabilities start being relevant already at large densities in the SN core, and are difficult to relate with the older treatments of slow instabilities, in which they were assumed to happen only \textit{outside} of the SN core (e.g. Refs.~\cite{Raffelt:2007cb,Duan:2009cd, Duan:2010bg, Mangano:2014zda, Duan:2014gfa, Abbar:2015mca,Abbar:2015fwa, Dasgupta:2015iia, Capozzi:2016oyk, Mirizzi:2015fva}), and were treated as a static flavor evolution along the direction of the neutrino stationary outflow. The SN surface acted in these early formulations as a stationary boundary (``bulb model'') from which perturbations would grow spatially. In fact, this formulation is part of a bigger disconnect between the older and the more recent literature on collective flavor conversions. In papers until approximately 2015, it was customary to consider the spatial evolution of neutrinos along their outflow direction; in more recent papers, concomitant with the discovery of the fast flavor instability, it has become customary to consider the temporal evolution of a system with a given initial condition (see, e.g., Refs.~\cite{Zaizen:2021wwl, Wu:2021uvt, Sigl:2021tmj, Shalgar:2022lvv, Richers:2022bkd, Bhattacharyya:2022eed}). The relation between the two formulations of the problem has been somewhat forgotten as a question. In general, it is not even clear whether a formulation in terms of an initial condition extracted from a SN profile is a realistic or correct one at all. Ideally we would like to solve the time evolution of the neutrino gas simultaneously with the matter background in which it evolves, but since this is beyond our computational reach, the question remains: how do fast and slow instabilities evolve in space and time? Should their evolution be treated as a stationary amplification in space, or as a temporal growth? 

The situation becomes physically more transparent in light of our recent interpretation of instabilities as resonant emission of flavor waves by neutrinos~\cite{Fiorillo:2024bzm,Fiorillo:2024uki,Fiorillo:2024pns}. As an instability first appears in the evolution of an astrophysical environment, wavepackets of flavor waves move with the group velocity, determined by the real part of the eigenfrequency, while they grow with a rate determined by its imaginary part. The flavor waves can be understood as a new, emergent degree of freedom, and as we have recently shown~\cite{Fiorillo:2025npi}, the energy of their quanta, the flavomons, can be uniquely defined as an independent degree of freedom. If the imaginary part is large enough, the wavepackets grow \textit{within} the region where they are first produced, and where the original perturbation is located. The instability is then said to be \textit{absolute}, inducing first a local relaxation. Instead, if the imaginary part is not large enough, the flavor wavepackets move so rapidly that they grow only away from the region where they are first produced. The instability is then defined \textit{convective}. In this case, a beam of neutrinos in a localized region exhibiting an instability steadily sources flavor waves which grow while leaving the region, leading to a spatial amplification rather than a temporal growth within the original site of emission. (The classification of instabilities as absolute or convective is standard in plasma physics and was introduced to the area of flavor conversion in Ref.~\cite{Capozzi:2017gqd} and also studied in Ref.~\cite{Yi:2019hrp}.) These arguments suggest that absolute instabilities lead to a local growth in time, while convective instabilities lead to a non-local amplification in space. On the other hand, these conclusions cannot be taken as too generic, and must always be adapted to the physical problem at hand.

\begin{figure}
    \includegraphics[width=\textwidth]{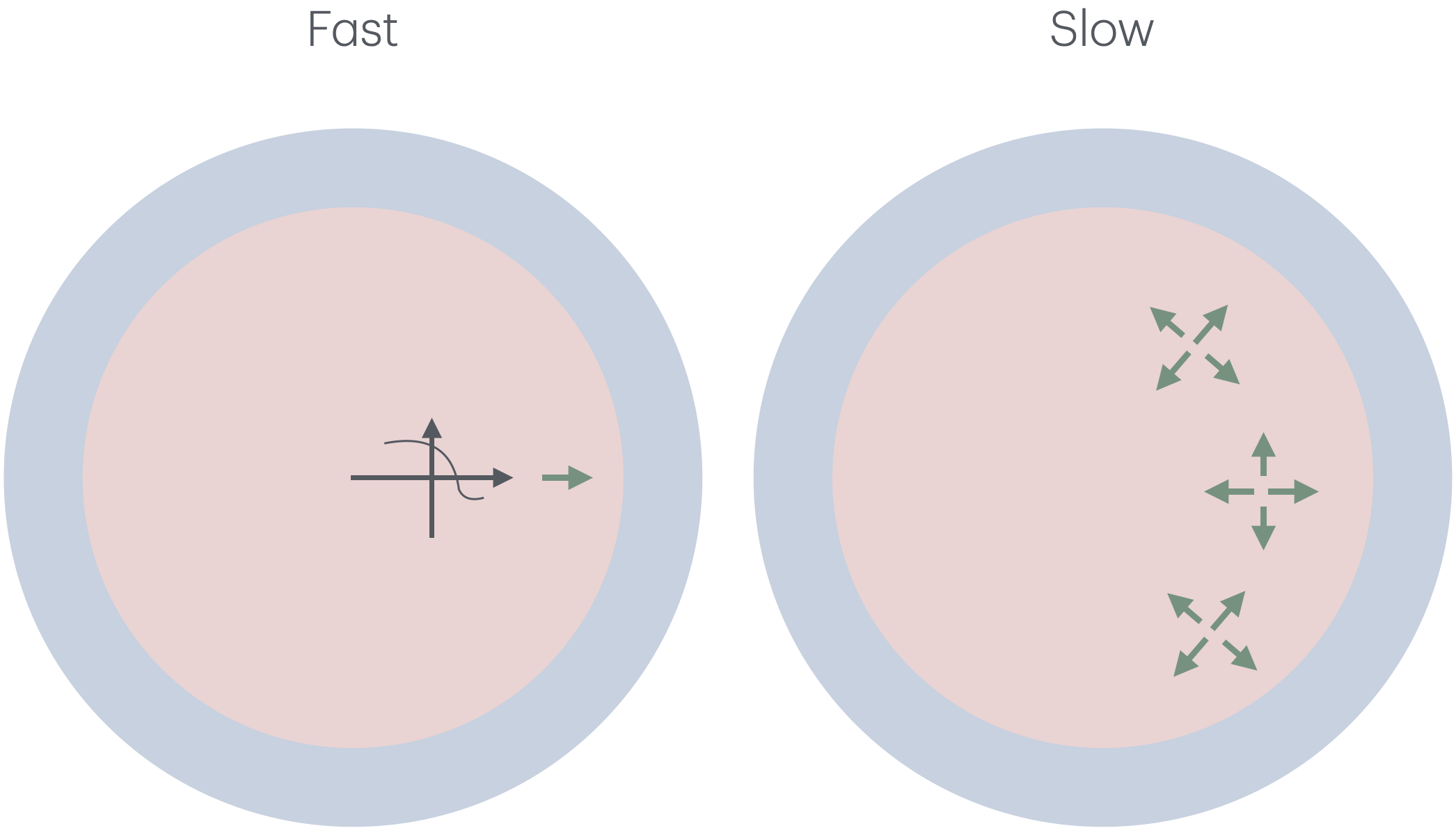}
    \caption{Pattern of instability growth for fast and slow instabilities. The green arrows show the direction along which a weak initial perturbation grows convectively. For fast instabilities, this direction depends on the neutrino angular distribution, sketched within the left panel; the small ``flipped'' side of the crossing -- similar to a beam of antineutrinos moving through a neutrino gas -- determines the direction of the growing modes. On the other hand, weak slow instabilities in an approximately isotropic medium are also convective, but without a preferred direction of motion.   
    }\label{fig:sketch}
\end{figure}

These simple physical insights must now be connected with the flavor instabilities identified for dense neutrino gases. Here we consider this question, with a special focus on the small-scale slow instabilities we have identified in Part~I. Our main conclusion is that all instabilities, both fast and slow, are convective when they are sufficiently weak. The resonance picture makes this result particularly intuitive: for a weak instability, flavor waves grow resonantly, so that they are preferentially emitted in phase with the ultra-relativistic neutrinos, therefore moving with the speed of light. Since in any physical system an instability must appear at first as a weak instability~\cite{Fiorillo:2024qbl,Fiorillo:2024bzm,Fiorillo:2024uki}, at the onset of instability one can never truly neglect the large-scale spatial structure of the neutrino flow. Besides identifying weak instabilities as convective, we also point out a fundamental difference between slow and fast instabilities. For fast instabilities, the growing waves are all directed towards the ``flipped'' region of the crossing, i.e., in the direction of the neutrinos carrying opposite lepton number to the bulk of the medium, similar to the sketch in Fig.~\ref{fig:sketch}. On the other hand, slow instabilities do not depend on the existence of a crossing, and therefore unstable waves can grow resonantly in all directions in which there are moving neutrinos, regardless of the lepton number they carry. Therefore, close to the decoupling region, where neutrinos are not too anisotropic, we expect that, while the instability is convective, growing waves move in all direction.

To show these results, we proceed in Sec.~\ref{sec:eom} to first summarize the equations of motion (EOM) that we use to describe collective conversions; here we also summarize the resonance picture of the instability that we developed in Refs.~\cite{Fiorillo:2024bzm, Fiorillo:2024uki, Fiorillo:2024pns}. In Sec.~\ref{sec:abs_conv} we briefly review the formal distinction between absolute and convective instabilities in a general sense. Finally, in Sec.~\ref{sec:neutrino_stream}, we consider a continuous angular distribution for neutrinos and determine the properties of fast and slow instabilities for such a system. Our primary interest is on slow instabilities within the core, which are primarily inhomogeneous as we have discussed in Part~I, but we also reconsider the problem of fast instability, showing that in the resonant picture their convective nature is easily understood. Finally, we summarize and discuss our results in Sec.~\ref{sec:discussion}.

\clearpage

\section{Dispersion relation and collective instabilities}\label{sec:eom}

\subsection{Quantum kinetic equations}

The flavor content of a dense neutrino gas is best described in terms of a density matrix $\varrho(\bp,\br,t)$ that depends on momentum $\bp$, space $\br$, and time $t$. The diagonal elements $\rho_{\alpha\alpha}=f_\alpha$ are the usual distribution functions for flavor $\alpha$, which in our two-flavor context is taken to be $\alpha=e$ or $\mu$. The off-diagonal elements, that we denote as $\psi_{\alpha\beta}=\rho_{\alpha\beta}$, represent the coherence between flavor $\alpha$ and $\beta$, and are non-vanishing if there are neutrinos in a superposition state between different flavors. In linear approximation, it is the flavor field $\psi_{\alpha\beta}$ that carries all the dynamics. In general, the evolution of this density matrix is determined by a set of quantum kinetic equations (QKE)~\cite{Dolgov:1980cq, Rudsky, Sigl:1993ctk, Sirera:1998ia, Yamada:2000za, Vlasenko:2013fja, Volpe:2013uxl, Serreau:2014cfa, Kartavtsev:2015eva, Fiorillo:2024fnl, Fiorillo:2024wej}
\begin{equation}
    (\partial_t+\bv\cdot\partial_\br)\varrho(\bp,\br,t)=-i[\sH(\bp,\br,t),\varrho(\bp,\br,t)],
\end{equation}
and similarly for the antineutrino density matrix $\overline{\varrho}(\bp,\br,t)$. We neglect any collisional term either among neutrinos or with matter, which could lead to new forms of flavor conversions and collisional instabilities~\cite{Johns:2021qby,Xiong:2022zqz, Liu:2023pjw, Lin:2022dek, Johns:2022yqy, Padilla-Gay:2022wck, Fiorillo:2023ajs}. The refractive Hamiltonian felt by neutrinos is
\begin{equation}\label{eq:Hamiltonian}
    \sH(\bp,\br,t)=\pm\frac{\sM^2}{2E}+\sqrt{2}\GF\sN
    +\sqrt{2}\GF\int\frac{d^3\bp'}{(2\pi)^3}\,\bigl[\varrho(\bp',\br,t)-\overline\varrho(\bp',\br,t)\bigr](1-\bv'\cdot\bv),
\end{equation}
where $\sM$ is the neutrino mass matrix and the negative sign applies to antineutrinos. In the rest frame of the medium, matter refraction is determined by the matrix $\sN$ of net charged fermion densities, i.e., it has $n_{e^-}-n_{e^+}$ etc.\ on the diagonal. The last term in Eq.~\eqref{eq:Hamiltonian} corresponds to refraction induced by other neutrinos, and is therefore the self-interaction term driving collective neutrino flavor evolution.

We assume that the evolution proceeds in axial symmetry along a coordinate that we call $r$ and denote by $v=\cos\theta$ the cosine of the angle between that direction and the neutrino velocity $\bv=\bp/|\bp|$, leading to the dimensionally reduced equation
\begin{equation}\label{eq:boltzmann}
    \left(\partial_t+v\partial_r\right)\varrho(v,r,t)=-i[\sH(v,r,t),\varrho(v,r,t)].
\end{equation}
The density matrices $\varrho(v,r,t)$ are now normalized to ${\rm Tr}\int_{-1}^{+1}dv\,\varrho(v,r,t)=n_\nu/(n_\nu+n_{\overline{\nu}})$, the relative local number density of neutrinos of all flavors, so that they are dimensionless. With this choice, the effective Hamiltonian becomes 
\begin{equation}
    \sH(v,r,t)=\pm\frac{\sM^2}{2E}
    +\sqrt{2}\GF\sN+\mu \int_{-1}^{+1}dv'\,\bigl[\varrho(v',r,t)-\bar\varrho(v',r,t)\bigr](1-v'v),
\end{equation}
where the effective neutrino-neutrino interaction strength is \hbox{$\mu=\sqrt{2}\GF (n_\nu+n_{\overline{\nu}})$}. There is an implicit assumption that $\varrho$ depends on space only through the coordinate $r$. We are therefore neglecting the possibility of spontaneous breaking of homogeneity in the plane transverse to that direction. 

One commonly proceeds by assuming that neutrinos in a certain region of space, or initially at a certain moment in time, are very close to flavor eigenstates, so that $|\psi_{\alpha\beta}|\ll f_\alpha$. Which of the two formulations should be taken as more realistic is essentially the topic of this paper; regardless of this question, close to this region in space or time, the equations can be linearized in terms of $\psi_{\alpha\beta}$. The linearized equations have been obtained in this form in Paper~I~\cite{Fiorillo:2024pns}, so we only briefly recall the main steps. Assuming only two flavors, any Hermitian $2\times 2$ matrix $\sA$ can be expressed in terms of a polarization vector $(A_0,\vec{A})$ as $\sA=\frac{1}{2}(A_0\sigma_0+\vec{A}\cdot\vec{\sigma})$. Here, $\vec{\sigma}$ are the Pauli matrices, $\sigma_0$ is the $2{\times}2$ unit matrix, $A_0={\rm Tr}\,\sA$, and $A_i={\rm Tr}(\sA\sigma_i)$ with $i=1$, 2 or 3. In the flavor basis, the matter term is written as $\frac{1}{2}\lambda\sigma_3$ with $\lambda=\sqrt{2}\GF(n_{e^-}-n_{e^+})$ if only electrons and positrons are present. Moreover, we write the vacuum oscillation term in the usual convention and notation
\begin{equation}
    \frac{\sM^2}{2E}=\frac{m_2^2+m_1^2}{2E}\frac{\sigma_0}{2}
    +\frac{m_2^2-m_1^2}{2E}\,\frac{\vec{B}\cdot\vec{\sigma}}{2},
\end{equation}
with $m_1<m_2$ being the masses of the two neutrino mass eigenstates. Here the ``magnetic field'' is a unit vector in flavor space which in the flavor basis has the components
\begin{equation}
    \vec{B}=(\sin2\theta_{\rm V},0,-\cos2\theta_{\rm V}),
\end{equation}
where $\theta_{\rm V}$ is the vacuum mixing angle. The vacuum oscillation frequency is denoted by
\begin{equation}
    \omegaE=\left|\frac{m_1^2-m_2^2}{2E}\right|,
\end{equation}
defined to be positive. Thus, the Hamiltonian describing vacuum flavor evolution is
\begin{equation}
    {\sf H}_{\rm V}=\frac{\omegaE}{2}
    \begin{pmatrix}-\cos2\theta_{\rm V}& \sin2\theta_{\rm V}\\
    \sin2\theta_{\rm V}&\cos2\theta_{\rm V}\end{pmatrix}.
\end{equation}
Since $\omegaE$ is positive, $\cos2\theta_{\rm V}>0$ implies normal mass ordering, while $\cos2\theta_{\rm V}<0$ implies inverted ordering. 

In order to proceed with the linearization, we introduce the polarization vectors describing the density matrices of neutrinos $\vec{P}=\mathrm{Tr}(\varrho\vec{\sigma})$ and antineutrinos $\vec{\bar{P}}=\mathrm{Tr}(\overline{\varrho}\vec{\sigma})$, and their difference $\vec{D}=\vec{P}-\vec{\bar{P}}$. The small quantities describing flavor coherence can now be described by $\psi=P^x+iP^y$ and $\overline{\psi}=\overline{P}^x+i\overline{P}^y$, while we denote the $z$ components simply by $P=P^z$ and $\overline{P}=\overline{P}^z$ and $D=P-\overline{P}$. The evolution equations for $\psi$ and $\overline{\psi}$, already linearized, are now
\begin{subequations}\label{eq:eom_linearized_inhomogeneous}
    \begin{eqnarray}
        \kern-2.5em
        (\partial_t+v\partial_r)\psi&=&
        -i(\omega_E \cV-\lambda)\psi+i\mu\bigl[\,\psi (D_0-v D_1)-P (\Psi_0-v \Psi_1)\bigr] -i\omega_E \sV P,
        \\
        \kern-2.5em
        (\partial_t+v\partial_r)\opsi&=&
        +i(\omega_E\cV+\lambda)\opsi +i\mu\bigl[\,\opsi (D_0-v D_1)-\oP (\Psi_0-v \Psi_1)\bigr] +i\omega_E \sV\oP,
    \end{eqnarray}
\end{subequations}
where we have introduced the notation $\Psi(v)=\psi(v)-\overline{\psi}(v)$ and the moments of the distributions $D_n=\int dv v^n D(v)$ and similarly for $\Psi$. The notation $\cV=\cos2\tV$ and $\sV=\sin2\tV$ is used for compactness. Here $\lambda=\sqrt{2}\GF (n_{e^-}-n_{e^+})$ accounts for matter refraction. In this work, we assume monoenergetic neutrinos and antineutrinos, otherwise in the moments $D_n$ and $\Psi_n$ we should also integrate over energies.

The terms proportional to $\sV$ act here as a source term, since they are the only inhomogeneous terms which do not vanish as $\psi$ and $\opsi$ tend to zero. If the homogeneous part of this system admits exponentially growing solutions, then the system is unstable and allows for runaway modes of flavor conversions. The key question of LSA is therefore to diagnose the possible existence of such unstable solutions for the homogeneous part. The strategy to verifying the existence of unstable solutions is usually to consider a volume element which is small compared to the characteristic scale over which a SN or NSM profile changes, so that we may take it to be homogeneous. We then look for solutions $\psi\to\psi e^{iKr-i\Omega t}$, the normal modes of the system. The approximation of homogeneous background medium only holds if $K^{-1}$, the wavelength of the modes, is much smaller than the scale of inhomogeneity of the environment. Therefore, this entire approach to LSA can only diagnose the existence of unstable small-scale modes. 

As we now summarize, both in fast and slow flavor conversions, the primary unstable modes are inhomogeneous with a very short wavelength, retrospectively justifying this approximation. With this form for the solution, and introducing the usual shifted variables $\omega=\Omega+ \mu D_0+\lambda$ and $k=K+ \mu D_1$, we find
\begin{subequations}\label{eq:longitudinal_eigenmode}
    \begin{eqnarray}
        \psi&=&\frac{P}{\omega-k v -\tomegaE}(\Psi_0-v \Psi_1),
        \\
        \opsi&=&\frac{\oP}{\omega-k v +\tomegaE}(\Psi_0-v \Psi_1),
    \end{eqnarray}
\end{subequations}
with $\tomegaE=\omegaE\cV$. Here we have also set $\mu=1$, with a suitable redefinition of the units of time and space. So far, we have implicitly assumed that $\psi$ depends only on $v$, and has no dependence on the direction of the momentum in the plane transverse to the spatial coordinate $r$. In Part~I, we have explored also the axially-breaking modes in which $\psi$ depends on this additional direction, whereas here, we focus on the longitudinal modes only. By requiring consistency of Eq.~\eqref{eq:longitudinal_eigenmode}, we obtain the dispersion relation
\begin{equation}\label{eq:dispersion_relation}
    (\tilde{I}_0-1)(\tilde{I}_2+1)-\tilde{I}_1^2=0,
\end{equation}
where
\begin{equation}\label{eq:tilde_I_n}
    \tilde{I}_n=\int \frac{P(v) v^n}{\omega-kv-\tomegaE}dv-\int \frac{\oP(v) v^n}{\omega-kv+\tomegaE}dv.
\end{equation}
These integrals are ill-defined for subluminal modes, i.e., with $|(\omega\pm \tomegaE)/k|<1$, due to the singularity in the denominator $\omega-kv\pm \tomegaE=0$. 

Indeed, there are no normal modes with $\omega$ real and subluminal and which are collective; instead, the normal subluminal modes with real eigenfrequency are the Case--van Kampen modes, which we introduced in the context of fast flavor conversions in Refs.~\cite{Fiorillo:2023mze,Fiorillo:2024bzm} and for slow flavor conversions in Part~I, and which correspond to non-collective modes streaming with individual neutrinos. Instead, below the light cone there can be a collective form of damping, 
known as Landau damping, corresponding to flavor waves decohering among the continuum of Case--van Kampen modes. The Landau-damped modes are not normal modes, but they correspond to physical excitations of the medium, in contrast to individual Case--van Kampen modes that have singular wave functions and as such are not physical. To capture the Landau-damped modes, the definition of the integrals $\tilde{I}_n$ needs be modified~to
\begin{equation}\label{eq:I_n}
    I_n=\int \frac{P(v) v^n}{\omega-kv-\tomegaE+i\epsilon}dv-\int \frac{\oP(v) v^n}{\omega-kv+\tomegaE+i\epsilon}dv,
\end{equation}
where the $i\epsilon$ prescription is enforced by causality and it schematically indicates that the integral over $v$ must be done over a contour in the complex plane passing below the singularity in the denominator; so if $\mathrm{Im}(\omega)<0$ the contour must be deformed to pass below the singular point. The unstable modes are captured by either form of the dispersion relation; however, because the unstable modes are analytically the continuation of the Landau-damped modes, using the integrals $I_n$ in place of $\tilde{I}_n$ allows us to find all of the modes, since in this analytical version of the dispersion relation, modes can only appear and disappear on the light cone, but are otherwise continuous functions. 

\subsection{Summary of collective instability branches}

Let us briefly review what is currently known about the unstable solutions of Eqs.~\eqref{eq:eom_linearized_inhomogeneous}. In the limit $\omega_E\to 0$, the system may admit unstable solutions, the fast flavor instabilities we have already discussed. In this case, we easily see that Eqs.~\eqref{eq:longitudinal_eigenmode} effectively depend only on the difference $\Psi=\psi-\opsi$, and we can write
\begin{equation}\label{eq:normal_fast_mode}
    \Psi=\frac{D}{\omega-kv}(\Psi_0-v\Psi_1).
\end{equation}
The existence of unstable solutions with $\mathrm{Im}(\omega)>0$, first noted by Sawyer~\cite{Sawyer:2004ai, Sawyer:2008zs, Sawyer:2015dsa,Chakraborty:2016lct}, was recognized to be relevant to a SN environment in Ref.~\cite{Izaguirre:2016gsx} and boosted the entire research on flavor conversions. While these early works have been primarily driven by the existence of unstable solutions to the mathematical system, in Refs.~\cite{Fiorillo:2024bzm, Fiorillo:2024uki} we have developed a theory of neutrino fast flavor conversions that makes their appearance intuitive. In this viewpoint, the four-vector $\Psi^\mu=(\Psi^0,\Psi^1,0,0)$ is regarded as a field, similar to the electromagnetic field; the elementary excitations of this field are usually called \textit{flavor waves}. A neutrino with velocity $v$ can absorb or emit flavor waves, according to Eq.~\eqref{eq:normal_fast_mode}, with an amplitude proportional to $D(v)$.
A flavor wave whose frequency satisfies the resonance condition $\omega=kv$ for some neutrinos moving with $\cos\theta=v$ can be emitted or absorbed by that neutrino in a Cherenkov process; this is already signaled by the vanishing denominator in Eq.~\eqref{eq:normal_fast_mode}. In the absence of an angular crossing, the spontaneous growth of flavor waves is impossible due to lepton number conservation~\cite{Johns:2024bob, Fiorillo:2024bzm}, so the only process allowed is Cherenkov absorption; the flavor waves are Landau-damped. In the presence of an angular crossing, a ``flipped'' region appears in which $D(v)$ changes sign. If there are flavor waves resonant with neutrinos in this region, the Cherenkov absorption turns into Cherenkov emission; waves are not Landau-damped, but rather compulsively emitted by neutrinos, leading to the exponential growth of the collective field $\Psi^\mu$.

In Part~I of this series, we have shown that, even in the absence of angular crossings, the system of Eqs.~\eqref{eq:longitudinal_eigenmode} admits unstable solutions if $P(v)$ and $\oP(v)$ have equal sign. The existence of slow instabilities, once the mass splitting $\tomegaE$ are nonzero, has long been known. However, we have shown that the properties of slow instabilities are dramatically different depending on the relative asymmetry between $P(v)$ and $\oP(v)$. If $P(v)$ and $\oP(v)$ are very close together for all velocities -- we roughly denote by $\epsilon=\mathrm{max}_v\left[(P-\oP)/(P+\oP)\right]$, so that the concrete condition can be formulated as $\epsilon\ll \sqrt{\tomegaE/\mu}$ -- then the properties of the instability are similar to the historical slow flavor pendulum. Unstable modes grow with a rate $\gamma=\mathrm{Im}(\omega)\sim \sqrt{\tomegaE \mu}$, and have wavenumbers which can reach up to $k\sim \sqrt{\tomegaE \mu}$. On the other hand, within a SN core, the ratio $\sqrt{\tomegaE/\mu}\sim 10^{-3}-10^{-2}$, so unless the neutrino and antineutrino fluxes are equal up to the percent level, we probably are not in this regime. 

Instead, in the opposite regime $\epsilon \gg \sqrt{\tomegaE/\mu}$, we have shown that slow instabilities still exist, but with very different properties. The typical growth rates are much slower, of the order $\gamma\sim\tomegaE/\epsilon$, while only modes with large wavenumber, of the order of $K\sim \mu \epsilon$, can become unstable. These instabilities are very close to the light cone, meaning that their phase velocity $\omega/k$ is very close to $1$ in any direction. Presumably this is the most relevant form of slow instability within a SN core or a NSM, where $\epsilon$ may be expected to be of the order of $10\%$, although a concrete study based on specific simulation models has not yet been performed.

\subsection{Source terms of the instability}\label{sec:source_terms}

For the purpose of determining the space-time evolution of an instability, we will often refer to the source term of the instability, i.e., the initial seed of flavor coherence which grows exponentially. In particular, it is of primary importance to know what is the space-time distribution of this source term. A detailed discussion of this topic is lacking from the literature, in which the instability is often said to be triggered by the vacuum mixing angle -- or sometimes by an effective in-matter mixing angle -- without clarifying what this entails for the space-time dependence of the source term. 

In the context of the EOM for collective flavor evolution, what drives the instability is the term in Eqs.~\eqref{eq:eom_linearized_inhomogeneous} proportional to $\sV$. Clearly this is proportional to the neutrino and antineutrino density $P$ and $\overline{P}$. Since we are assuming a homogeneous background medium, at first sight it would appear that the source term is only present for the homogeneous mode $K=0$. However, in reality the neutrino density will never be perfectly homogeneous, and will always possess small-scale fluctuations of small amplitude. Such fluctuations are unavoidably driven by the small inhomogeneities of the matter profile which in turn are driven e.g.\ by hydrodynamic turbulence. Hence, terms driving inhomogeneous modes are always present, albeit with amplitudes much smaller than the term driving the homogeneous mode. 

Nevertheless, the former are presumably much more important, at least at the onset of the instability. First of all, what we call the homogeneous mode -- which has been studied in detail and is often associated with regular pendulum-like behavior -- is actually ill-defined in a realistic environment, since its wavelength is comparable with the large-scale variations of the SN or NSM profile. So the approximation of infinite medium underlying the dispersion relation breaks down completely. Secondly, even if the seed for homogeneous or large-scale modes with $K\sim 0$ is very large, the instability grows only if these modes are unstable. As we will see, when an instability first appears, it is very weak, and the range of unstable wavenumbers is very narrow; for fast instabilities, this has been shown in Refs.~\cite{Fiorillo:2024bzm,Fiorillo:2024uki,Fiorillo:2024dik}, while for slow instabilities in Part~I of this series. It would be quite coincidential if a narrow interval of unstable wavenumbers which are of the order of $K\sim \mu\epsilon$ would contain exactly the value $K=0$. Without fine tuning, one expects only small-scale modes to be unstable at the onset of instability, and therefore the seed for small-scale modes, albeit being much smaller, is the most important driving term in the EOM.

Therefore, in a realistic environment we expect the seed term to be proportional to $\omega_E \sV \delta P(r,t)$, where $\delta P(r,t)$ is the fluctuating component of the neutrino density, and analogously for antineutrinos. We have added here also a time dependence; generally, if the initial neutrino configuration at $t=0$ is $\delta P^0(r)$, at a time $t$ due to the streaming of neutrinos this will be $\delta P(r,t)=\delta P^0(r-vt)$. The situation is of course even more complex, because in reality neutrinos are still collisionally coupled to the medium, which fluctuates also in time due to hydrodynamic turbulence, which means that $\delta P(r,t)$ will likely fluctuate in time and space following the medium, but to capture the latter effect, the collisional coupling should be taken into account. For the purpose of this work, we can schematically consider the source term to be essentially free; so when we discuss a localized perturbation, we can schematically represent it as a localized ``bump'' $\delta P^0(r)$ which propagates with velocity $v$. It is clear that in reality this source term will appear at any point in space and time, with consequences to be discussed later.

\section{Classification of instabilities}\label{sec:abs_conv}

\subsection{Absolute and convective instabilities}

In the field of fast flavor evolution, a distinction between absolute and collective instabilities was drawn early \cite{Capozzi:2017gqd, Yi:2019hrp}. While these works present a formal discussion, elaborating on the criteria that separate the two cases, an intuitive explanation of their nature (e.g.\ why fast instabilities can transition from convective to absolute as they become stronger), and especially a discussion of what are the practical consequences of this distinction, is still missing. Here we will revise this question, both for fast and slow flavor evolution.

We begin with a short recap of the concept of absolute and convective instabilities. For an unstable system, a plane-wave perturbation with a wavevector $K$ can grow exponentially. However, realistic perturbations are not in the form of plane waves, but rather in the form of localized structures. Therefore, a localized perturbation may propagate so rapidly that it grows only far away from its original location. This possibility was first pointed out by Twiss~\cite{twiss1951oscillations, twiss1959growth, twiss1952propagation} and Landau and Lifshitz~\cite{landau1987fluid}. In the linear phase of evolution, such structures can be written as a superposition of plane waves, each oscillating and potentially growing if its eigenfrequency $\omega(k)$ has a positive imaginary part. 

A simple heuristic argument to differentiate between absolute and convective instabilities can be made following Kadomtsev~\cite{kadomtsev1965plasma}. We assume that the growth rate is maximum at a wavenumber $k_0$, so that around this value we may expand
\begin{equation}\label{eq:expansion_dispersion_relation}
    \omega\simeq \omega_0+U \delta k+\frac{\alpha}{2}\delta k^2+i\left(\gamma_0-\frac{\beta}{2}\delta k^2\right),
\end{equation}
where $\delta k=k-k_0$, whereas $U$, $\alpha$, and $\beta$ are coefficients. For simplicity, we consider that the collective field is described by a single quantity, say, $\Psi_0$; it is easy to generalize the argument to any number of degrees of freedom. The evolution in time of $\Psi_0$ is therefore given by
\begin{equation}
    \Psi_0(r,t)=\int dk\,e^{i\left[kr-\omega(k) t\right]}\tilde{\Psi}_0(k,0),
\end{equation}
where $\tilde{\Psi}_0$ is the Fourier transform of the initial wavepacket. One can reasonably expect that after a while only amplitudes for wavenumbers around $k_0$ will be relatively large. Therefore, we may evaluate $\tilde{\Psi}_0(k,0)\simeq \tilde{\Psi}_0(k_0,0)$. The remaining integral yields explicitly
\begin{eqnarray}
    \Psi_0(r,t)&\simeq&\tilde{\Psi}_0(k_0,0) \sqrt{\frac{2\pi}{(\beta+i\alpha)t}}
    \nonumber\\[1.5ex]
    &&\kern2em{}\times
    \exp\left[i\left(k_0 r-\omega_0 t+\frac{\alpha(x-U t)^2}{2t(\alpha^2+\beta^2)}\right)+\left(\gamma_0t-\frac{\beta(x-Ut)^2}{2t(\alpha^2+\beta^2)}\right)\right].
\end{eqnarray}
At any fixed position, the perturbation initially grows, but as $t\to \infty$, it will keep growing only if
\begin{equation}\label{eq:critical_velocity}
    U^2<\frac{2\gamma_0(\alpha^2+\beta^2)}{\beta}.
\end{equation}
Therefore, there is a critical group velocity for the most unstable mode, above which the unstable wavepacket at any fixed position actually \textit{is damped} with time. The simple reason is that the wavepacket propagates so rapidly that it convects away the instability which hence is called convective. Vice versa, if at any point in space the perturbation ultimately grows unbounded, the instability is absolute. 

This reasoning is only heuristic because it uses an expansion of the dispersion relation close to the wavenumber carrying maximum growth. A more formal approach~\cite{Sturrock:1958zz} requires one to consider the evolution of $\Psi_0(r,t)$ for an arbitrary dispersion relation at infinitely late times, leading to a simple, albeit somewhat formal, criterion of discrimination. The instability is absolute if there exists a (generally complex) value of $k$ for which $d\omega/dk=0$ and $\mathrm{Im}(\omega)>0$, i.e., there must be an unstable eigenmode with vanishing group velocity. Intuitively, such an eigenmode with vanishing group velocity ``stays'' in the system while growing. In fact, we can easily see that Eq.~\eqref{eq:critical_velocity} corresponds exactly to the condition that $\mathrm{Im}(\omega)>0$ for $d\omega/dk=0$ with the dispersion relation Eq.~\eqref{eq:expansion_dispersion_relation}. In other words, Eq.~\eqref{eq:critical_velocity} is correct provided that the mode with vanishing group velocity is sufficiently close to the one with maximum growth rate.

If a system exhibits an absolute instability, there is no question that its evolution must be formulated as an evolution in time starting from a given initial configuration. On the other hand, the situation is more puzzling if there are only convective instabilities. Since the instability is convective, any initial perturbation from a region centered around a point $r_0$ will grow while moving away, effectively disappearing. If the source of the perturbation remains active at $r_0$ -- for our collective instabilities this is more or less ensured, because the source term discussed in Sec.~\ref{sec:source_terms} remains non-vanishing in proportion to $\delta P(r_0,t)$ -- the instability is continuously sourced. In the immediate vicinity of $r_0$ the perturbation then remains small, but it grows spatially away from $r_0$, due to the overlapping growing wavepackets emitted from $r_0$. One might therefore be tempted to treat this as a boundary-value problem, with the boundary condition of a vanishing perturbation at $r_0$ and the source term discussed in Sec.~\ref{sec:source_terms}. Therefore, one might conclude that if the dispersion relation admits solutions with $\mathrm{Im}(k)<0$ for some real $\omega_0$, and the perturbation is sourced at a frequency $\omega_0$, there would be spatial amplification of the perturbations for $r>r_0$, so that the factor $e^{ik(r-r_0)}$ grows exponentially.

However, while in many circumstances this strategy may provide the correct answer, one should remember that this is \textit{not} in general a correct procedure. The existence of a solution with complex $k(\omega_0)$ alone does not ensure its appearance in a concrete physical problem. As a simple counter example, consider a system of two neutrino beams (no antineutrinos) with the same number of electron neutrinos moving with opposite velocities; so we take $v_1=1$, $v_2=-1$, and $P_1=P_2=1$. The integrals $I_n$ in Eq.~\eqref{eq:I_n} reduce to simple sums, and the final dispersion relation Eq.~\eqref{eq:dispersion_relation} reduces to
\begin{equation}
    \omega^2=k^2+4;
\end{equation}
we have taken $\tomegaE=0$ for simplicity, since there are no antineutrinos and it therefore would make no difference. Obviously there are no unstable modes because our system does not contain initially any antineutrino or muon neutrino, and so pairwise conversions are not possible. Nevertheless, for $-2<\omega<2$, the system admits modes with $\mathrm{Im}(k)\neq 0$, both positive and negative. Clearly these modes cannot correspond to a spatial growth, which is forbidden by lepton number conservation. 

We can easily understand what has gone wrong; since the two beams are moving in opposite directions, information flows both ways, preventing a boundary-value problem. This serves as a warning that general statements regarding the correct formulation of the problem are difficult to establish. Such statements can be made based on formal criteria, which were obtained long ago in plasma physics. In particular, one can show \cite{briggs1964electron} that spatially growing solutions exist if (i)~the system has only convective instabilities (if it has no instability, as in the two-beam example above, there can be no spatial growth; and if it has absolute instabilities, the perturbation grows locally and a boundary-value problem would be ill-defined anyway), and (ii)~if the system is excited at a frequency $\omega_0$ such that $\mathrm{Im}[k(\omega_0)]<0$ (for $r>r_0$) and $\mathrm{Im}[k(\omega_0+i M)]<0$ with $M\to \infty$. Such criteria are hard to apply in practice, and they do not particularly help to develop an intuitive understanding of the behavior of the instability. Therefore, rather than relying exclusively on these formal results, we will undertake a discussion of collective neutrino instabilities based primarily on specific examples, and use physical arguments to understand the absolute versus convective nature of the instability and what this entails in practice for its evolution.

\subsection[Analytical properties of the dispersion relation for complex \texorpdfstring{$k$}{}]{Analytical properties of the dispersion relation for complex \texorpdfstring{\boldmath$k$}{}}

Before proceeding with these practical examples, let us briefly reconsider how the usual dispersion relation must be extended to include complex values of $k$. This is a mostly technical intermezzo that is not directly needed to follow our later physical discussion, but is important to reproduce our results for the solutions with complex $k$.

For a system with discrete degrees of freedom, e.g., a discrete system of neutrino beams, the dispersion relation is a polynomial in $\omega$ and $k$. Therefore, its analytical continuation to complex values of these variables is simple. For a continuous angular distribution, however, the situation becomes considerably more complicated. We have already discussed in multiple papers~\cite{Fiorillo:2023mze, Fiorillo:2024bzm, Fiorillo:2024uki, Fiorillo:2024pns, Fiorillo:2024dik} how the analytical continuation must be performed for real $k$ and complex $\omega$. To ensure causality, any integral of the form
\begin{equation}\label{eq:definition, positive_omega}
    F(\omega,k)=\int_{-1}^{+1}\frac{f(v)\,dv}{\omega-kv},
\end{equation}
where $f(v)$ is an arbitrary function of the velocity (the discussion is trivially generalized if $\omega\to \omega\pm \tomegaE$ in the denominator), must be extended with the $i\epsilon$ prescription
\begin{equation}\label{eq:Landau_definition}
    F(\omega,k)=\int_{-1}^{+1}\frac{f(v)\,dv}{\omega-kv+i\epsilon}.
\end{equation}
Formally, this means that the analytical continuation must be done by continuously lowering the imaginary part of $\omega$ from $\mathrm{Im}(\omega)\to \infty$ to $\mathrm{Im}(\omega)=\omega_I$, the required value. Since for $\mathrm{Im}(\omega)\to+\infty$, the pole of the integrand in Eq.~\eqref{eq:Landau_definition} is at $v=\omega/k$, so $\mathrm{Im}(v)\to\pm\infty$ for $k>0$ or $k<0$ respectively, this means that the integration contour in $v$ must be deformed to pass \textit{below} the pole $v=\omega/k$ if $k>0$, or \textit{above} if $k<0$. Hence, we can maintain the original definition in Eq.~\eqref{eq:definition, positive_omega} for $\mathrm{Im}(\omega)>0$. In the opposite case, we must deform the contour to surround the singularity, which introduces a separate contribution to the integral coming from the small circle surrounding the singularity
\begin{equation}
    F(\omega,k)=\int_{-1}^{+1}\frac{f(v)\,dv}{\omega-kv}-\frac{2\pi i}{|k|}f\left(\frac{\omega}{k}\right)\quad\hbox{for}\quad
    \mathrm{Im}(\omega)<0.
\end{equation}
An additional consequence is that the hard integration boundary at $v=\pm 1$, if $\mathrm{Im}(\omega)<0$ and $\mathrm{Re}(\omega/k)=\pm 1$, makes the integral ill-defined. The nature of this singularity is a branch cut~\cite{Fiorillo:2024bzm}, and so $F(\omega,k)$ for real $k$ has two branch cuts at $\mathrm{Re}(\omega)=\pm k$ and $\mathrm{Im}(\omega)<0$.

We must now extend this definition to complex values of $k$. This extension is defined once again by the general principle that for $\mathrm{Im}(\omega)\to \infty$, the integral must be defined as in Eq.~\eqref{eq:definition, positive_omega}, and as $\mathrm{Im}(\omega)$ is lowered to finite values, the integration contour in $v$ must be deformed such as to never touch the singularity $v=\omega/k$. If we write $\omega=\omega_R+i\omega_I$ and $k=k_R+ik_I$, this singularity lies at
\begin{equation}
    v=\frac{(\omega_R k_R-\omega_I k_I)+i(\omega_I k_R-\omega_R k_I)}{k_R^2+k_I^2}.
\end{equation}
As $\omega_I\to \infty$, this pole lies \textit{above} the real axis of $v$ for $k_R>0$, and \textit{below} it for $k_R<0$. Provided that $\omega_I-\omega_R k_I/k_R>0$, the pole remains above (or below) the real axis of $v$, so there is no need to deform the contour. Therefore, for generic values of $\omega$ and $k$, the integral remains defined as in Eq.~\eqref{eq:definition, positive_omega} if $\omega_I-\omega_R k_I/k_R>0$. In the opposite case, we must deform the contour so that
\begin{equation}
    F(\omega,k)=\int_{-1}^{+1}\frac{f(v)\,dv}{\omega-kv}-\frac{2\pi i}{k\; \mathrm{sign}(k_R)}f\left(\frac{\omega}{k}\right)
    \quad\hbox{for}\quad
    \omega_I-\omega_R k_I/k_R<0
\end{equation}
after acquiring once again a contribution from the small circle surrounding the pole.

We have thus found the general definition for the integral. It ensures analyticity of $F(\omega,k)$, except at the branch cuts. However, we can now realize that these are always in the half space $\omega_I-\omega_R k_I/k_R<0$, but they correspond to $\mathrm{Re}(v)=\pm 1$, i.e., to the lines
\begin{equation}
    \frac{\omega_R k_R+\omega_I k_I}{k_R^2+k_I^2}=\pm 1.
\end{equation}
These are the generalized branch cuts of $F(\omega,k)$, which is particularly important information, because these are the places in the complex plane where the solutions of the dispersion relation may disappear abruptly, whereas otherwise they must vary continuously. Later we will seek solutions with real $\omega$ and complex $k$, so having identified these branch cuts is an important guide, since we now know that solutions with complex $k$ and real $\omega$ can disappear abruptly if
\begin{equation}\label{eq:branch_cut_dispersion}
    \frac{\omega k_I}{k_R}>0
    \quad\hbox{and}\quad
    \frac{\omega k_R}{k_R^2+k_I^2}=\pm 1.
\end{equation}
Our examples will indeed show that complex branches of $k$ disappear abruptly if this condition is verified.

\section{Space-time evolution of collective instabilities}\label{sec:neutrino_stream}

In this section, we discuss the absolute versus convective nature of the main forms of collective instabilities in the flavor evolution of a dense neutrino gas. Our primary aim is to clarify that even the formulation of the problem, i.e., its initial or boundary condition, may be more subtle than had been assumed.

\subsection{Stable configuration}\label{sec:stable}

\begin{figure}
    \includegraphics[width=\textwidth]{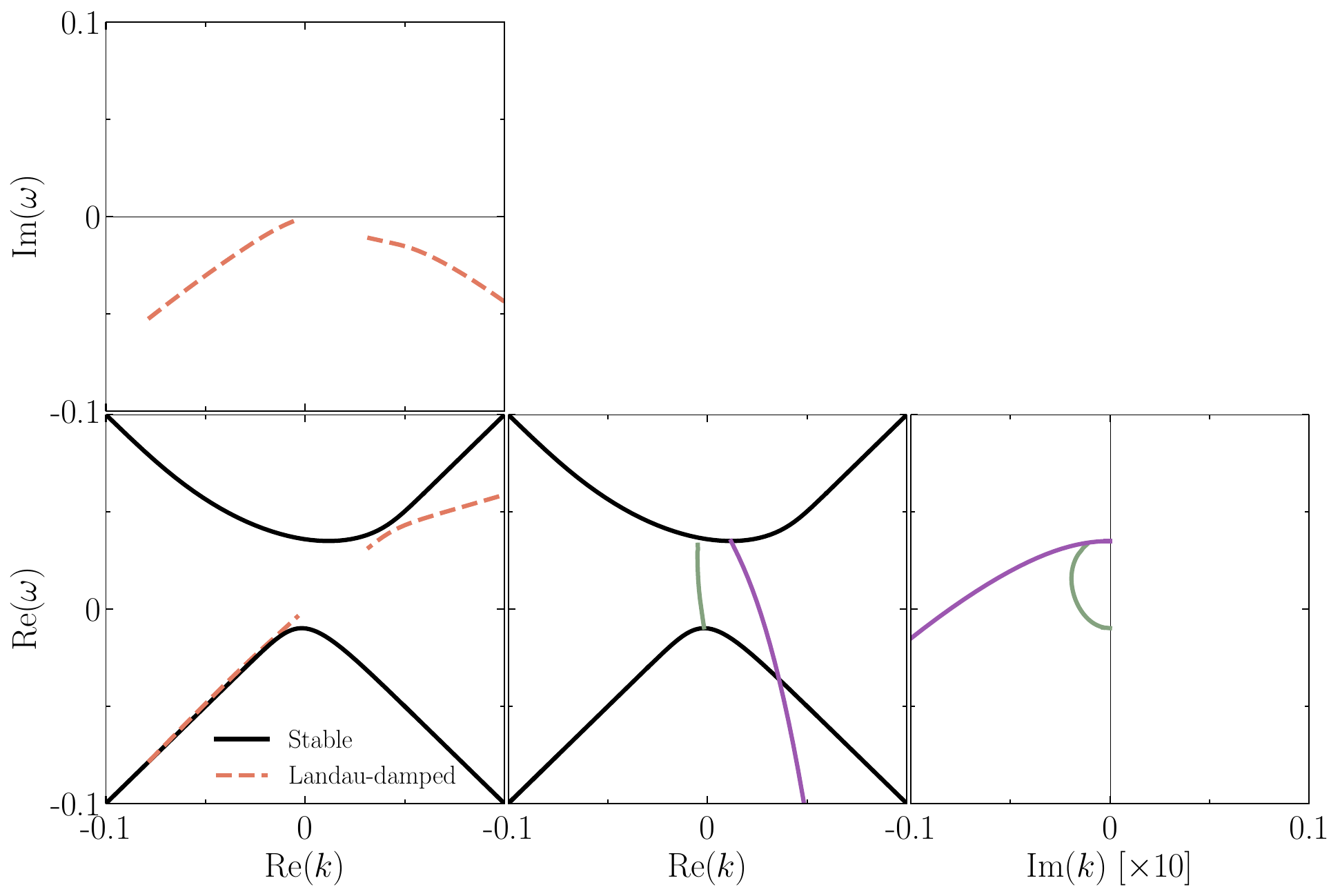}
    \caption{Solutions of the dispersion relation for the stable angular distribution of Eq.~\eqref{eq:benchmark}. {\it Left panels:} complex $\omega$ and real $k$. {\it Center and right panels:} real $\omega$ and complex $k$, in black the branches with $\mathrm{Im}(k)=0$, in colors those
    with $\mathrm{Im}(k)\neq0$.}\label{fig:stable}
\end{figure}

As a first case, we begin with a completely stable example, which already reveals nontrivial questions concerning the dispersion relation. To this end, we use an uncrossed angular distribution in the fast flavor limit (no vacuum frequency). Specifically, we use the same angular distribution previously adopted in Paper~I
\cite{Fiorillo:2024pns} and shown there in Fig.~1
\begin{equation}\label{eq:benchmark}
    P(v)=0.5
    \quad\hbox{and}\quad
    \oP(v)=0.47+0.025\exp\left[-(1-v)^2\right].
\end{equation}
There is no angular crossing and therefore no fast instability and there is no vacuum frequency and therefore no slow instability either. In Fig.~\ref{fig:stable}, we show the solutions of the dispersion relation Eq.~\eqref{eq:dispersion_relation} for the physical modes, i.e., using the modified definition of the integrals $I_n$ in Eq.~\eqref{eq:I_n}. We show both the solutions with complex $\omega$ and real $k$ (left panel), and the solutions with real $\omega$ and complex $k$ (center and right panels).

In the left panels, there are no unstable branches, but there are two branches of Landau-damped modes (dashed lines) already familiar from Paper~I. The damping is understood as decoherence over the continuum of Case--van Kampen modes or, in a different language, as Cherenkov absorption of flavor waves on neutrinos resonant with the mode. More interestingly, in the center panel we see the emergence of branches of complex $k$ for real frequencies $\omega$. Such branches are also shown, for different angular distributions, in Ref.~\cite{Yi:2019hrp}, although there they disappear abruptly at $\omega=0$. With our modified dispersion relation, which has the correct analytical properties enforced by causality, we see that these branches do not disappear at $\omega=0$, but rather they continue until they reach a branch cut of the dispersion relation, which occurs where Eq.~\eqref{eq:branch_cut_dispersion} is satisfied.

What is the physical meaning of these branches of complex $k$? Since the system as a whole is stable (it does not possess modes growing in time), they cannot correspond to spatial amplification. While this is obvious from a physical perspective, we have verified it explicitly by checking that, for a given point on the branch of complex $k$, if we increase $\mathrm{Im}(\omega)$ continuously from $0$, the value of $\mathrm{Im}(k)$ never changes sign. As we discussed in Sec.~\ref{sec:abs_conv}, this ensures the absence of spatial amplification. Instead, these modes correspond to a regime of \textit{non-transparency}; if the system is steadily forced at a temporal frequency $\omega$ for which a mode with complex $k$ appears, the perturbation will gradually dissipate away from the region where the forcing happens. Physically this effect is similar to driving a plasma perturbation with a frequency below the plasma frequency---the perturbation exponentially declines with distance from the source.

\subsection{Fast instability}\label{sec:fast}

\begin{figure*}
    \includegraphics[width=\textwidth]{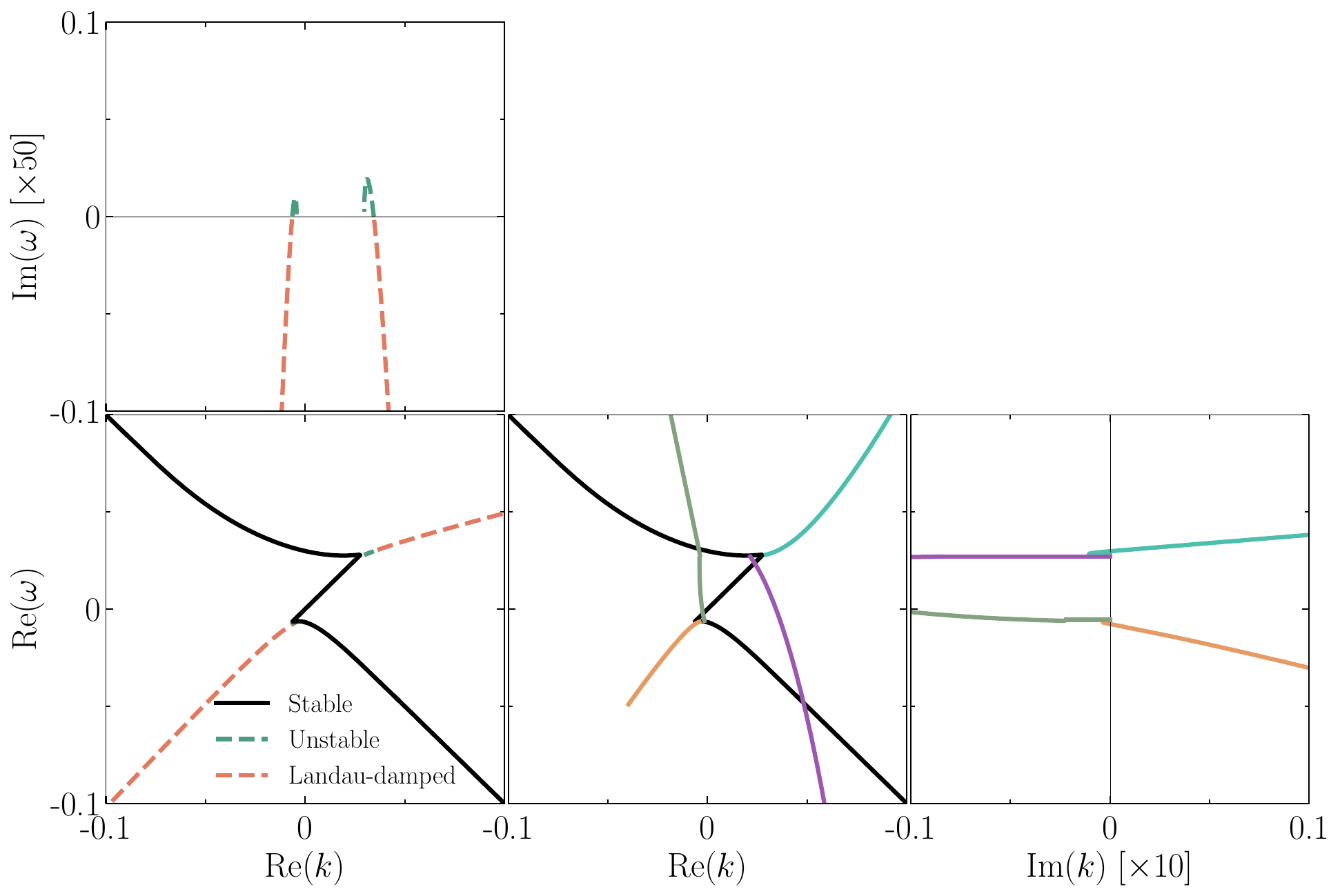}
    \caption{Same as Fig.~\ref{fig:stable}, now for the crossed angular distribution of Eq.~\eqref{eq:benchmark_fast}.}\label{fig:crossed}
\end{figure*}

We now slightly deform the angular distribution so that an angular crossing emerges, while maintaining $\tomegaE=0$, implying a fast instability. The general properties of the dispersion relation were first studied in Ref.~\cite{Yi:2019hrp}, and more recently in Ref.~\cite{Fiorillo:2024dik} in light of the more appropriate analytical dispersion relation introduced in Refs.~\cite{Fiorillo:2024bzm, Fiorillo:2024uki}. Here we focus on the case of a very shallow angular crossing, following the general principle that at its onset an instability is always weak~\cite{Fiorillo:2024qbl}. Therefore, we use as an angular distribution
\begin{equation}\label{eq:benchmark_fast}
    P(v)=0.5
    \quad\hbox{and}\quad
    \oP(v)=0.47+0.0302\exp\left[-(1-v)^2\right];
\end{equation}
the specific numerical values are chosen to maximize visibility of the branches with complex wavenumbers.

Figure~\ref{fig:crossed} shows the solutions of the dispersion relation with real $k$ and complex $\omega$ (left panel) and with real $\omega$ and complex $k$ (center and right panels). For real $k$, we observe the typical structure already found in Ref.~\cite{Fiorillo:2024dik}, i.e., the two stable branches of the uncrossed case (Fig.~\ref{fig:stable}) merge into a single branch that crosses the light cone. In turn, the Landau-damped branches, which previously were confined below the light cone with a negative growth rate, can now develop a positive growth rate when their phase velocity becomes larger than the crossing velocity $\vc$ where the lepton number $P(v)-\oP(v)$ changes sign. These unstable modes finally merge with the stable branch outside of the light cone; this merging happens at the points where the stable branch has a \textit{vertical} tangent, i.e., where $d\omega/dk\to +\infty$, as proven in Refs.~\cite{Yi:2019hrp, Fiorillo:2024uki}. 

We stress that for such weakly unstable angular distributions -- which are the physical cases of interest since an instability always appears first within a weakly unstable configuration -- the range of wavenumbers that becomes unstable is extremely narrow, and contains wavenumbers of the order $k\sim \mu \epsilon$. Since this range is so narrow, it would be coincidental or fine tuned if it happened to contain also the special value $k=G_1$ or $K=0$, the homogeneous mode. Therefore, we conclude that at their onset fast instabilities are always small-scale, and for a weakly unstable angular distribution the homogeneous mode is stable unless fine-tuned cases are considered.

The first question is whether the instability is absolute or convective, and the answer first provided in Ref.~\cite{Yi:2019hrp} is that it is convective. To see this explicitly, we use the insight from Sec.~\ref{sec:abs_conv} that an instability is absolute if there is a mode with generically complex $k$ and $\omega$, vanishing group velocity $d\omega/dk=0$, and $\mathrm{Im}(\omega)>0$. In this case, we see that the stable branch, with $\mathrm{Im}(\omega)=0$, has two pairs of points with horizontal tangent and vanishing group velocity: one in the upper half-plane, and one in the lower. These pairs of modes are stable, and therefore cannot lead to a local growth of the instability. Hence, the instability as a whole is convective. As the angular crossing becomes deeper, each pair of points with horizontal tangent can disappear, turning the instability absolute~\cite{Yi:2019hrp,Fiorillo:2024dik}. Here, we do not explore this regime, and focus on this single example of a very shallow angular crossing.

The convective nature of the instability can be viewed from another perspective. The unstable modes appear only in the narrow vicinity of the crossing velocity $\vc$, which for a weak crossing is very close to $1$. Therefore, these modes have, at their onset, both a phase and a group velocity very close to the speed of light. This is not a purely mathematical property; it descends from the resonance picture of the fast instability~\cite{Fiorillo:2024bzm,Fiorillo:2024uki}, so that the growing modes are the ones that can be resonantly emitted from the neutrinos beyond the crossing. Therefore, the unstable modes move with a preferential direction (we might say ``towards'' the crossing) with a velocity very close to the speed of light. 

Since the instability is convective, an initial localized perturbation moves away from that region, in a direction defined by the angular crossing. Therefore, there must be spatial amplification in this direction and one may ask about this growth rate. Therefore, we now focus on the branches of complex wavenumber $k$ for real frequency $\omega$. They appear in correspondence to every point at which the real branch has a horizontal tangent $d\omega/dk=0$, in the same way that branches of complex $\omega$ appear where the real branch has a vertical tangent. Therefore, there are four branches of complex $k$. Two of them -- the green and purple ones in Fig.~\ref{fig:crossed} -- are only a slight deformation of the branches with the same colors already existing for the stable case in Fig.~\ref{fig:stable}. Therefore, these branches do not correspond to a spatial amplification, but to a non-transparency of the medium. Instead, the cyan and orange branches in Fig.~\ref{fig:crossed} appear only because of the crossing, and therefore must lead to spatial amplification. Indeed, from their imaginary parts (right panel of Fig.~\ref{fig:crossed}), we see that in a very narrow range of frequencies close to the onset of the mode they are negative. This is the range of frequencies for which spatial amplification appears. We can prove this formally by considering how the imaginary part $\mathrm{Im}(k)$ for these solutions changes by increasing $\mathrm{Im}(\omega)$; we have explicitly tested that $\mathrm{Im}(k)$ changes sign in these narrow frequency intervals, confirming that they lead to spatial amplification.

However, it is simpler to understand this result based on our physical picture. The unstable modes move with the speed of light and with a positive velocity, i.e., to the right. Therefore, these two branches can lead to a spatial amplification only if $\mathrm{Im}(k)<0$, so that $e^{ikz}\propto e^{|\mathrm{Im}(k)|z}$ grows in the direction of the unstable modes. If a perturbation has frequency components within the narrow range where the cyan and orange curves have $\mathrm{Im}(k)<0$, such components will exponentially grow in space, rather than in time. We should also note that the point at which the curves for $\mathrm{Im}(k)=0$ correspond to a phase velocity equal to the crossing velocity $\vc$, and in fact they intersect the branches of complex $\omega$ and real $k$ where they also pass through $\mathrm{Im}(\omega)=0$, transitioning from unstable to Landau-damped. This allows us to identify rather easily the extremal frequencies at which the spatial amplification appears: they simply correspond to $\mathrm{\omega}_{\pm}=\vc k_\pm$, where $k_\pm$ are the wavenumbers marking the appearance of (subluminal) unstable modes, which we determined in Ref.~\cite{Fiorillo:2024uki}. The frequency range of spatial amplification has also two superluminal boundaries, where the complex $k$ branches merge with the real branches, which correspond to the superluminal boundaries of the range of unstable wavenumbers $k$; as discussed in our previous works, these threshold values cannot be determined by a simple explicit formula.

In summary, we have shown that a weak fast instability has a preferential direction, i.e., the one towards the shallow crossing region. Unstable modes grow in this direction, and therefore a stationary source of perturbations can produce a spatially growing signal in that direction, provided that the perturbations contain Fourier frequencies in the narrow range in which $\mathrm{Im}(k)<0$ for the branches passing through the crossing velocity.

\subsection{Slow instability in a near-isotropic gas}\label{sec:slow_near_isotropic}

\begin{figure}
    \includegraphics[width=\textwidth]{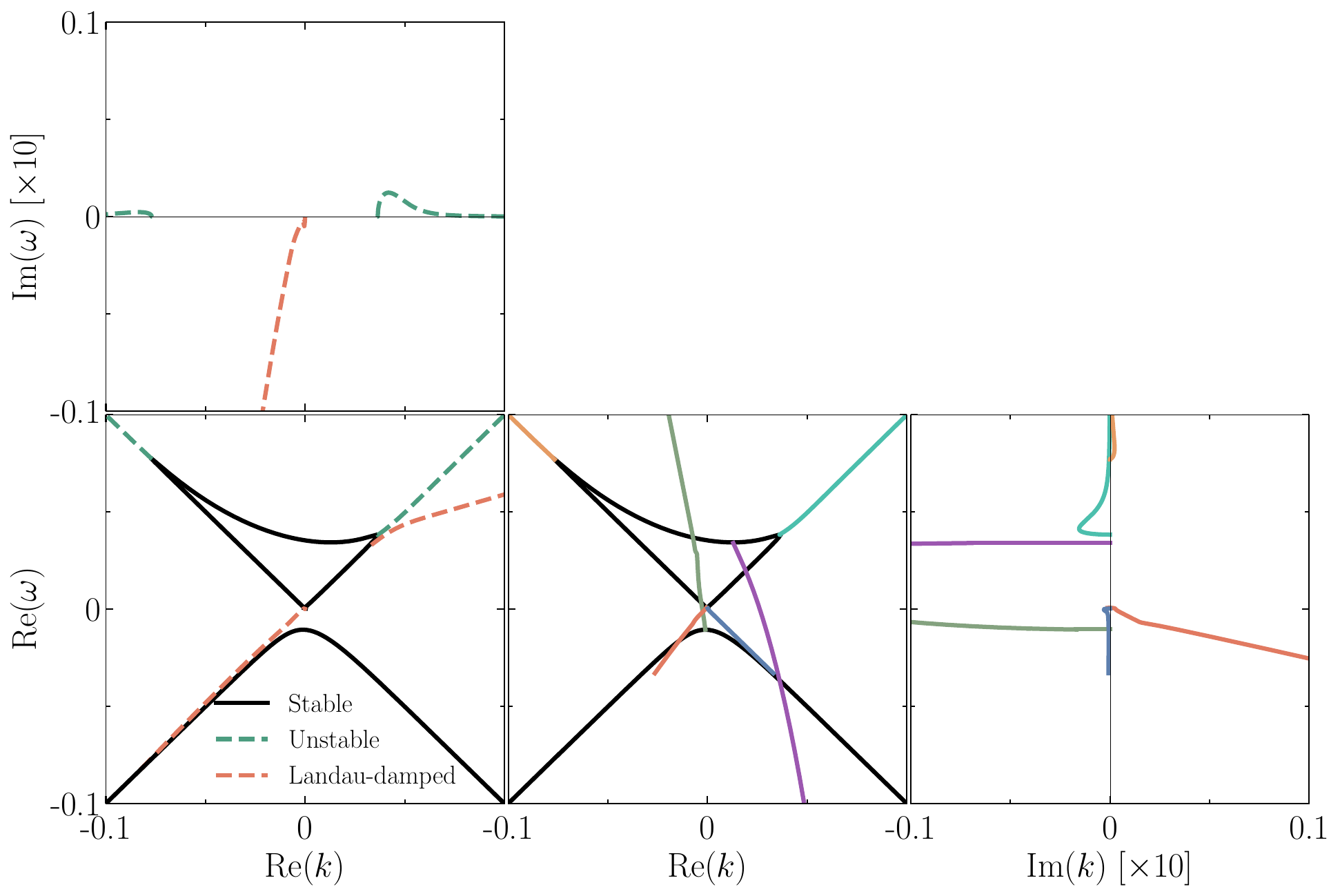}
    \vskip-12pt
    \caption{Same as Fig.~\ref{fig:stable} with the same uncrossed angular distribution of Eq.~\eqref{eq:benchmark}, but now with a vacuum oscillation frequency $\tomegaE=-10^{-5}$, leading to a slow instability of the resonant type.}\label{fig:slow}
\vskip12pt
    \includegraphics[width=\textwidth]{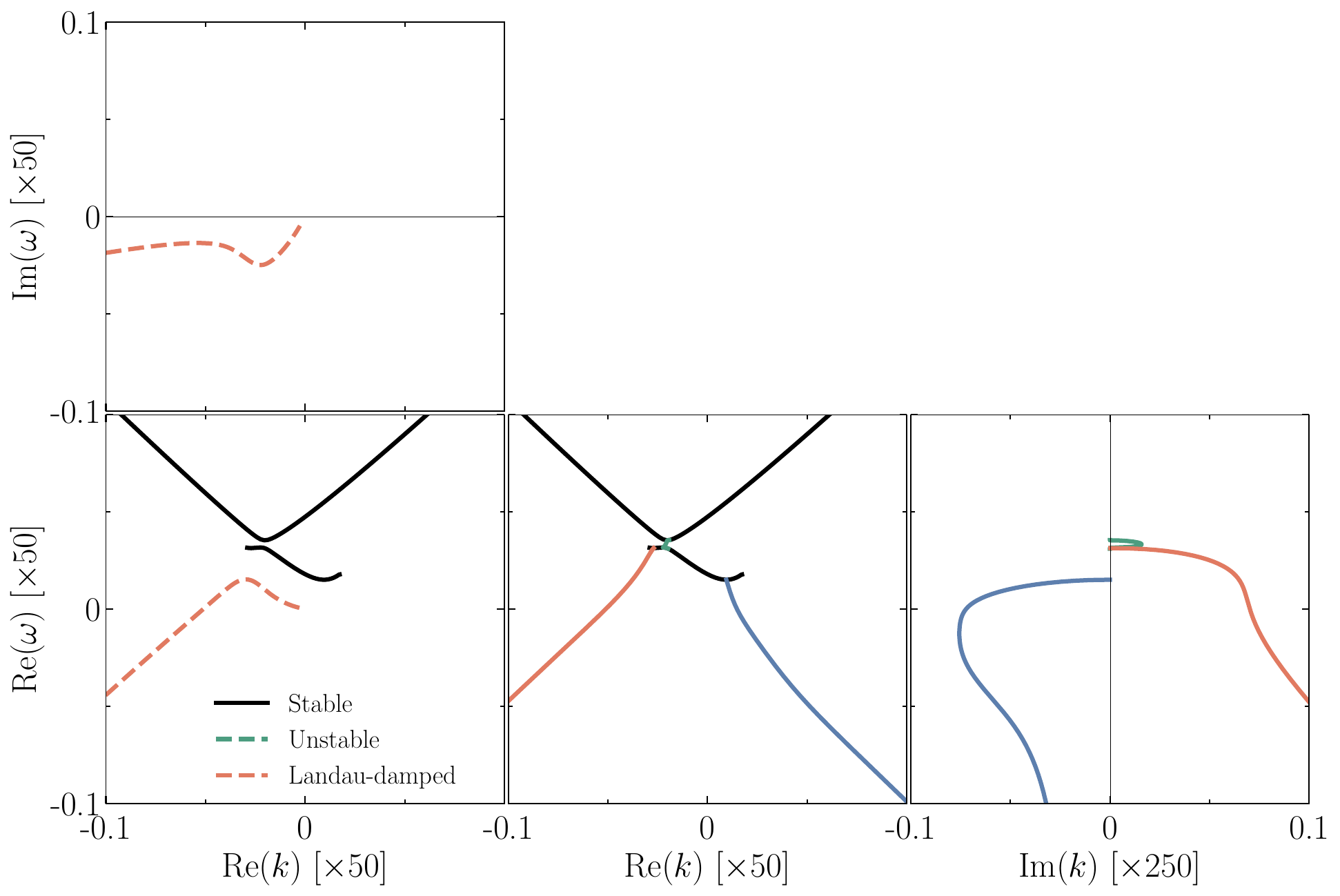}
    \vskip-12pt
    \caption{Zoomed-in version of Fig.~\ref{fig:slow}, to highlight the novel modes appearing close to the origin.}\label{fig:slow_zoom}
\end{figure}

We next modify the original stable case of Sec.~\ref{sec:stable} in a different way, i.e., by introducing a non-vanishing vacuum frequency. In Paper~I, we have systematically discussed the different regimes as the vacuum frequency is increased, both in normal and inverted ordering. Here we focus on a single case, namely inverted ordering with $\tomegaE=-10^{-5}$, whereas the angular distribution is again the uncrossed one of Eq.~\eqref{eq:benchmark}. We are therefore looking at a case where $|\tomegaE|\ll \mu \epsilon^2$, which in Paper~I we have identified as the \textit{resonant} regime.

The structure of the modes is shown in Fig.~\ref{fig:slow}. In the left panel, we show the unstable modes for real $k$ and complex $\omega$, which we have already analyzed in Paper~I \cite{Fiorillo:2024pns} (see Fig.~2 for comparison). To briefly review the structure of these modes here, note that at large $k$, what was a single stable branch for $\tomegaE=0$ (see Fig.~\ref{fig:stable}) produces two unstable branches moving close to the light cone. They merge into a real branch at smaller values of $k$. The stable branch close to the light cone for negative $\omega$ does not suffer strong changes due to the non-vanishing $\tomegaE$.

We again conclude that the slow instability is convective. Mathematically, we see that the real branch for $\omega>0$ in Fig.~\ref{fig:slow} has developed four points with horizontal tangent $d\omega/dk=0$, which therefore correspond to stable modes with vanishing group velocity. A more physical way to understand the convective nature of the instability is to note that once again the unstable waves feed on the resonant interaction with neutrinos moving along the light cone, and therefore the unstable waves have a group velocity close to the speed of light. This is easy to show mathematically; in Paper~I, we have shown that for the resonant modes, an approximate form for the frequency is $\omega=\pm k+\chi(k)$, with $\chi(k)\ll k$. Moreover, we have derived explicit expressions for $\chi(k)$ when $k$ is very large; for our purposes, we simply notice that since $\chi \ll k$, it follows immediately that $d\omega/dk\simeq \pm 1$. 

Since the instability is convective, we now ask about the spatial amplification it leads to, studying the structure of the modes for real $\omega$ and complex $k$. Once again, the green and purple branches are just slight deformations of the analogous ones appearing in the stable case in Fig.~\ref{fig:stable}, and therefore lead only to non-transparency. Instead, we now observe the appearance of two branches of complex $k$, the orange and cyan ones in the central and right panels of Fig.~\ref{fig:slow}. These branches merge with the real one at the points where the latter has a horizontal tangent $d\omega/dk=0$, and clearly correspond to the branches of complex $\omega$ and real $k$ in the left panel. In fact, we can show this explicitly; since the unstable branches are written as $\omega=k+\chi(k)$ with $\chi(k)\ll 1$, it follows immediately that there must also be solutions of the form $k=\omega-\chi(\omega)$ to first order in $\chi$, which therefore have a complex wavenumber and real frequency. Thus, the cyan and orange branches must correspond to spatial amplification, originating from the unstable wavepackets that are continuously sourced and leave their region steadily. Indeed, we see that the cyan branch, which moves with a positive group velocity has $\mathrm{Im}(k)<0$, so it grows for $z>0$, while the orange branch with a negative group velocity has $\mathrm{Im}(k)>0$, so it grows for $z<0$.

For very small $k$, there are additional branches of complex $k$; to see them clearly, we also show in Fig.~\ref{fig:slow_zoom} a zoomed-in version of Fig.~\ref{fig:slow}. The left panel of this figure was already shown in Paper~I. We now see that the two real branches formed in this very narrow region close to the origin are in fact connected by a branch of complex $k$ (shown in green), and two additional branches of complex $k$ (shown in blue and red) also originate from this region. Since these branches do not correspond to any branch of unstable frequency, we can already anticipate that they cannot lead to spatial amplification. Indeed, we have explicitly verified that for such solutions $\mathrm{Im}(k)$ does not change sign as $\mathrm{Im}(\omega)\to +\infty$, so they must correspond to non-transparency.

We conclude that the most interesting effect induced by slow instabilities corresponds to the spatial amplification of modes moving with the speed of light (the cyan and orange complex $k$ branches in this example). Any flavor perturbation grows exponentially with distance from its initial appearance, with a typical growth rate of the order of $\mathrm{Im}(k)\propto \tomegaE/\epsilon$. We have here an apparent paradox: while the spatial growth rate is very small, the real part $\mathrm{Re}(k)\sim \mu \epsilon$. The two quantities are however disconnected; the real part $\mathrm{Re}(k)\sim \mathrm{Re}(\omega)$ corresponds to the wavelength of the growing modes that leave the region, while their spatial growth rates measure the typical distance over which they become large. To consistently capture the growth of these slow modes, a spatial resolution comparable with $[\mathrm{Re}(k)]^{-1}\sim (\mu\epsilon)^{-1}$ is required.

Finally, from this discussion it might appear that slow modes move preferentially along both the positive and negative $z$ axis, as if this was a special direction in the system. In reality, however, as already discussed in Paper~I, this conclusion derives from limiting our discussion to modes with wavevector $\bk$ along the axis of symmetry. However, it is clear that there are unstable modes moving in any direction in which there are neutrinos; in Paper~I, we have shown that the typical growth rate of a mode along a given direction is proportional to $\mathrm{Im}(\chi)\propto \tomegaE [P(v)+\oP(v)]/[P(v)-\oP(v)]$, see Eq.~(5.19) in \cite{Fiorillo:2024pns}. Therefore, the space-time evolution of slow instabilities can be schematically represented as in Fig.~\ref{fig:sketch}: from a given region, there are unstable modes leaving in any direction in which there are neutrinos, and their growth rate is larger for the direction in which the ratio between lepton and neutrino number is smallest. Of course, if there are directions where this ratio becomes anomalously small due to an angular crossing, our approximations fail, but the reason is quite simple; along such directions, there will be fast-unstable modes which are then much more important than the slow modes we consider here.

\subsection{Slow instability far from the decoupling region}

We finally mention a topic mainly of historical relevance, namely the case of slow instabilities starting far from the decoupling region, where the neutrino angular distribution is strongly forward peaked. This was the primary concept of slow neutrino flavor instabilities in works before 2015, which were treated with a formalism characteristically different than the current one; rather than considering an initial condition for the unperturbed neutrino flow, which is then evolved in time, one took the opposite viewpoint of imposing a boundary condition on a region close to the central protoneutron star (neutrino bulb model) and evolving the flow in space for radii beyond this boundary~\cite{Duan:2009cd, Duan:2010bg}.

Our current work retrospectively justifies this attitude for the specific problem considered in these papers, and at the same time shows the limitations of this approach. As we have seen, the slow instability is convective, with unstable modes resonantly growing along any direction in which there are neutrinos. In a neutrino gas with a strongly forward-peaked angular distribution, if there are no neutrinos moving backward, this means that the spatial growth is only along the radial direction. Therefore, there will be a branch of real frequencies such that a small perturbation at the inner region, containing Fourier components at such frequencies, is spatially amplified at larger radii. Thus, the convective nature of the instability justifies the old ``spatial'' approach, showing however that the spatially growing signal needs not be perfectly static; there will be an entire range of frequencies whose modes are spatially amplified, a point first noted by Refs.~\cite{Dasgupta:2015iia, Abbar:2015fwa}. In fact, the convective nature of a forward-moving neutrino gas has greater generality than the slow instabilities only. In such a setup, since neutrinos can only move forward, the information flow also must only be forward, meaning that there can be no information about the growth of flavor perturbations propagating downstream of the flow. This statement, while intuitive, can be formally proven by noting that the EOM admit a set of characteristics along the neutrino trajectories, and that a fixed boundary at $r=\rm const.$ spawns only characteristics that leave the surface.

One aspect that was long discussed for slow instabilities in the SN context was the relevance of the so-called halo effect~\cite{Cherry:2012zw, Sarikas:2012vb}, i.e., a backward-reflected neutrino component which is small (perhaps 1\% of the forward-moving neutrinos), but with a wide angular distribution. Since the angular distribution contains in this case backward-moving neutrinos, the general arguments in favor of a boundary formulation of the problem fail, since there is a flow of information towards the inner PNS, as well as unstable slow modes moving along the direction of ingoing neutrinos. In fact, there might even be fast-unstable modes along such directions, if the halo particles are dominated by antineutrinos~\cite{Abbar:2021lmm,Zaizen:2021wwl}. However, if the amount of lepton number in the halo flux is sufficiently small, their impact is similarly small and does not alter directly the above statements. 

The halo neutrinos can be treated perturbatively in that their density matrix is regarded as a small perturbation, propagating in the unperturbed medium of the forward-moving neutrinos. Due to neutrino-neutrino interactions, the halo neutrinos can therefore exchange lepton number with the forward beam; this is indeed what happens in fast instabilities, where close to the crossing velocity there is a resonant growth of flavor waves which drain lepton number from the ``flipped'' region -- in this case, from the halo neutrinos -- carrying it to neighboring velocities -- in this case, into the forward-moving neutrinos. This is the main feedback that the halo neutrinos have on the forward-moving ones. For slow resonant instabilities, the impact is even more limited, since, as we have seen, unstable modes resonate only with neutrinos along the direction of their velocity, and therefore they can drain the halo neutrinos of their lepton number independently of neutrinos moving in other directions -- lepton number is here exchanged between neutrinos and antineutrinos moving collinearly, rather than between neutrinos moving across neighboring directions. Therefore, provided that the lepton number in the halo is much smaller than the one in the forward beam, the impact of the halo is perturbative and does not alter our previous statements. Ref.~\cite{Abbar:2015mca} reached a similar conclusion based on a numerical approach.

After having justified the historical treatment for the formulation in terms of slow instabilities far from the decoupling region, we can now finally understand why such treatment fails: the problem itself is ill-formulated. Slow instabilities can appear already much deeper, close to the decoupling region, even in the absence of angular crossings. In this region, the neutrino angular distribution is much less forward-peaked, and therefore unstable modes will move along all directions as sketched in Fig.~\ref{fig:sketch}. Therefore, a boundary formulation is ultimately unjustified due to the much earlier appearance of slow instabilities, close to the decoupling region. Understanding where such instabilities precisely arise is a topic that we will explore in an upcoming work.

\section{Discussion}\label{sec:discussion}

We have extended the theory of slow instabilities formulated in Paper~I \cite{Fiorillo:2024pns}, as well as the parallel theory of fast instabilities developed in Refs.~\cite{Fiorillo:2024bzm, Fiorillo:2024uki}, to answer an overlooked question: how does a \textit{localized} perturbation evolve in space and time? The growth rates of plane waves cannot provide the answer, because these are by definition not local.

Our main conclusions follow from the analogous theory of space-time evolution of plasma instabilities (e.g.\ Ref.~\cite{BersSpaceTime}). If the growing waves have too large a group velocity, they can escape the original region before growing, leading to a convective instability, while if their group velocity is not too large, some of them will grow locally, leading to an absolute instability. This classification was formally introduced to the flavor community in Refs.~\cite{Capozzi:2017gqd, Yi:2019hrp}, where the dispersion relation of fast-unstable systems was analyzed to determine whether their instability was absolute or convective. However, the question of what this entails for the space-time evolution of the instability was left open.

Our discussion was mainly guided by two key examples, relating to a fast-unstable and a slow-unstable system. In both cases, we focus on situations of weak instability, in which the growth rate is not very large; this is done in the fast-unstable case by considering a shallow angular crossing, while in the slow-unstable case by considering the limit $\tomegaE\ll \mu \epsilon^2$. The physical motivation for this choice, introduced in Refs.~\cite{Fiorillo:2024qbl,Fiorillo:2024bzm,Fiorillo:2024uki}, is that an instability, at its onset, is always weak, and before it can become very strong, the medium, coupled with flavor conversions, presumably reacts to avoid its formation, although this effect can only be captured by a self-consistent implementation of flavor conversions, feedback on the medium, and medium evolution.

In the fast-unstable case, a weak instability is convective. This result is not new \cite{Yi:2019hrp}, and we have recently re-discussed it in the context of the resonant theory of fast instabilities \cite{Fiorillo:2024dik}. Unstable waves are emitted in phase with neutrinos in the ``flipped'' region beyond the crossing, and hence have a group velocity close to the speed of light, moving together with these neutrinos and effectively leaving the system. This implies that the source of the perturbations, which ultimately is the mixing angle, produces a spatial growth in the direction of the flipped neutrinos, starting from the point where the crossing first appears. So, for example, if the crossing is produced by a beam of antineutrinos passing through a neutrino medium and starting at a point $z=0$, the intensity of the flavor field grows as $\propto e^{-\mathrm{Im}(k) z}$ spatially for $z>0$ if the flipped neutrinos move towards positive $z$. For stronger instabilities, with a more pronounced angular crossing, the instability becomes absolute and grows locally even within the region where the crossing is located~\cite{Fiorillo:2024dik}.

We may next ask what this means for a concrete formulation of the problem of fast instability. Many authors, based on numerical approaches, have considered small volume elements (periodic boxes) in which an initial perturbation grows in time; the medium is effectively treated as infinite and homogeneous. In reality,
a region containing a crossing is not infinite and thus, for a convective instability, flavor waves grow spatially from the boundaries. The range over which this growth happens is of the order of a few wavelengths $(\mu\epsilon)^{-1}$; since the size of the crossing region is comparable with the SN or NSM inhomogeneity scale $\ell\gg (\mu\epsilon)^{-1}$, numerical treatments based on periodic boxes can still describe flavor evolution at distances much larger than $(\mu\epsilon)^{-1}$ from the edge. Still, at sufficiently late times $\ell/v_{\rm gr}\sim \ell$, where $v_{\rm gr}$ is the group velocity of unstable waves, the effects of the boundaries will be felt anywhere within the unstable region, and the description in terms of periodic boxes will fail. 

For the slow-unstable case, that we have introduced in Paper~I, in the limit \hbox{$\tomegaE\ll \mu \epsilon^2$}, the instability is also formally convective. However, flavor waves now grow resonantly with neutrinos moving in all directions, different from the fast case where they only grow in a preferential direction, that of the flipped neutrinos. This means that perturbations grow spatially along any direction. In addition, even though the unstable slow modes have small scales, with a wavelength comparable to the fast modes of the order of $(\mu\epsilon)^{-1}$, they grow over much longer time scales, of the order of $\tomegaE^{-1}$. Since these scales are comparable with the light-crossing time over the inhomogeneity length scale of a SN profile $\ell$, a periodic-box approach would presumably not be helpful; within a time $\tomegaE$ in which the wave grows sizably, it has already traversed through the inhomogeneous medium over a length $\ell$. Notice that for slow instabilities, the more complete space-time treatment that we adopt here, rather than the usual plane-wave dispersion relation, is crucial, because the wavelengths and growth rates differ widely, so it is not immediately clear what should be compared with the inhomogeneity scale. The space-time approach we use here clarifies this point, showing that, even though the modes have a wavelength of the order of $(\mu\epsilon)^{-1}$, much shorter than the inhomogeneity scale $\ell$, they lead to spatial growth over length scales of order $\tomegaE^{-1}$, which is instead comparable.

In summary, we have argued that the often-used periodic boxes to describe fast instabilities likely cover only the early stages of instability growth. Boundary effects, and more generally, the impact of large-scale inhomogeneities, are crucial, especially for weak instabilities, because of their convective nature. For slow instabilities, the periodic-box approach seems even less likely to succeed, due to the comparable inhomogeneity length scale $\ell$ and spatial growth rate $\tomegaE^{-1}$. 

A more appropriate description might involve a full, large-scale inhomogeneous SN profile, and determining the flavor evolution therein based on the full quantum-kinetic equations, including collision terms. Such a formulation is followed in the works of the NBI group \cite{Shalgar:2022lvv, Shalgar:2022rjj, Cornelius:2023eop, Cornelius:2024zsb}, who have also stressed that the impact of a localized instability could be primarily felt elsewhere as a result of neutrino streaming. However, realistic implementations pose formidable challenges that have not yet been resolved. First of all, post-processed SN simulations, obtained \textit{without} accounting for flavor conversions, are usually adopted as a starting point, but these are intrinsically inconsistent, as we stressed in Ref.~\cite{Fiorillo:2024qbl}. Rather, one could start from simulations exhibiting weak instabilities, at their first appearance, although likely this would happen during the SN infall phase. As we have discussed, such instabilities only appear at very short scales, so they cannot be captured unless a sufficient spatial resolution, perhaps of the order of a few cm in the SN core, is achieved. For weak instabilities, a coarser resolution would completely overlook the existence of an instability altogether, but such a small numerical scale does not exist in practical SN simulations and therefore are entirely subgrid in numerical reality. 

Even for strongly unstable configurations, as they appear in the post-processing of inconsistent simulations, it remains unclear whether a coarse spatial resolution could capture the evolution of the instability, and if so, what would be the physical reason. From theory alone it would appear that a resolution comparable with the scale of a few cm would be needed in any case, since small-scale structure appears even in the early stages of linear evolution which can be captured by analytical studies. A recent numerical work also argues in this direction~\cite{Nagakura:2025brr}.

An additional limitation of current numerical treatments is that they should also include the usually neglected matter term, which cannot be eliminated in any obvious way, in contrast to what is sometimes stated, and whose role may be especially important due to its large-scale inhomogeneity. 

Since a complete treatment of these aspects seems computationally unfeasible, we argue that the periodic-box approach -- which as we have discussed is a good starting point for the \textit{early stages} of fast flavor evolution -- should be complemented by a study of the spatial evolution of unstable flavor waves in an inhomogeneous medium. This topic, which has not received much attention, seems crucial to determine the concrete evolution of flavor instabilities, both fast and slow, close to the decoupling region of a SN.

\acknowledgments

DFGF is supported by the Alexander von Humboldt Foundation (Germany),
whereas GGR acknowledges partial support by the German Research Foundation (DFG) through the Collaborative Research Centre ``Neutrinos and Dark Matter in Astro- and Particle Physics (NDM),'' Grant SFB-1258\,--\,283604770, and under Germany’s Excellence Strategy through the Cluster of Excellence ORIGINS EXC-2094-390783311.

\bibliographystyle{JHEP}
\bibliography{Biblio.bib}

\end{document}